\documentclass[11pt]{paper}
\usepackage{amssymb,latexsym, amsmath}
\usepackage{amscd}

\usepackage[dvips]{graphicx}

\newtheorem{exm}{Example}

\newtheorem{thm}{Theorem}

\newtheorem{as}{Assumption}

\newcommand{\R}{\mathbb{R}}

\newcommand{\so}{\mathfrak{so}}

\newcommand{\gl}{\mathfrak{gl}}

\DeclareMathOperator{\Ad}{\mathrm{Ad}}
\DeclareMathOperator{\Span}{\mathrm{span\,}}
\DeclareMathOperator{\diag}{\mathrm{diag}}

\DeclareMathOperator{\rank}{\mathrm{rank}}
 \DeclareMathOperator{\pr}{\mathrm{pr}}
\DeclareMathOperator{\tr}{\mathrm{tr}}

\begin{document}

\title{ON THE COMPLETENESS OF THE MANAKOV INTEGRALS}



\author{Vladimir Dragovi\' c $^1$,
 Borislav Gaji\' c $^2$ and Bo\v zidar
Jovanovi\'c $^3$}

\maketitle

\leftline{\small$^1$ Department of Mathematical Sciences,
University of Texas, } \leftline{$\,\,$ Dallas 800 West Campbell
Road 75080 Richardson TX, USA}
 \leftline{\small \footnote{\small E-mail:
Vladimir.Dragovic@utdallas.edu}\footnote{\small E-mail:
gajab@mi.sanu.ac.rs}\footnote{\small E-mail: bozaj@mi.sanu.ac.rs}
Mathematical Institute SANU, Kneza Mihaila 36, 11000 Belgrade,
Serbia }

\begin{abstract}
The aim of this note is to present  simple proofs of the
completeness of Manakov's integrals for a motion of a rigid body
fixed at a point in $\R^n$, as well as for geodesic flows on a
class of homogeneous spaces $SO(n)/SO(n_1)\times\dots\times
SO(n_r)$.
\end{abstract}

\

\centerline{\small\it Dedicated to  Academician Anatoly
Timofeevich Fomenko on the occasion of his 70th birthday.}

\

\section{Introduction}

In this note we consider a well known problem of a $n$-dimensional
rigid body motion, closely related to two important topics in the
theory of integrable systems: the construction of integrable
systems on Lie algebras and the conjecture that noncommutative
integrability implies the usual Liouville one by means of
integrals that belong to the same functional class as
noncommutative integrals. Both of these problems originated and
have been highly developed by Anatoly Timofeevich Fomenko and his
school \cite{MF1, MF2, TF, Bo}. Nowadays, the second problem is
known as the Mishchenko--Fomenko conjecture.

The Euler equations on $\so(n)$ of a motion of $n$--dimensional
rigid body around a fixed point are given by
\begin{equation}
\dot M_{ij}=\sum_{k=1}^n \frac{b_i-b_j}{(b_k+b_i)(b_k+b_j)}
M_{ik}M_{kj} \label{E3}
\end{equation}
(see \cite{Fr, Ra, FeKo}). Here $M\in\so(n)$ is the angular
momentum of the rigid body, related to the angular velocity
$\Omega$ by $M_{ij}=(b_i+b_j)\Omega_{ij}$, where
$B=\diag(b_1,\dots,b_n)$ is the mass tensor. They are Hamiltonian
with the Hamiltonian function being the kinetic energy of the body
$H=\frac12\sum_{i<j}M_{ij}\Omega_{ij}$, with respect to the
Lie-Poisson bracket
\begin{equation}
\{f,g\}(M)=-\langle M, [\nabla f(M),\nabla g(M)]\rangle, \qquad
M\in\so(n), \label{Lie-Poisson}
\end{equation}
where $\langle M_1,M_2\rangle=-\frac12\tr(M_1M_2)$, $M_1,M_2 \in
\so(n)$.

Following Dubrovin \cite{Dub, Du}, Manakov found the Lax
representation of the system and proved that the solutions of
(\ref{E3}) are expressible in terms of $\theta$-functions (see
\cite{Ma}). Another Lax pair of the system can be found in
\cite{Fe}.

Mishchenko and Fomenko in \cite{MF1} proved that the Manakov
integrals form a complete Poisson-commutative set
\begin{equation}
\mathcal L=\{\tr(M+\lambda A)^k\, \vert\, \lambda\in \mathbb{R},
\,k=1,2,\dots,n \}, \qquad A=B^2. \label{integrals}
\end{equation}

\begin{thm}\label{prva} Suppose that the matrix $A$ has distinct eigen-values.
Then,  there exist
\begin{equation}
\label{uslov0}
N_{\so(n)}=\frac12(\dim\so(n)+\rank\so(n))=\frac12\left(\frac{n(n-1)}{2}+\left[\frac{n}{2}\right]\right)
\end{equation}
independent polynomials in $\mathcal L$. The set $\mathcal L$ is a
complete Poisson-commutative set on $\so(n)$. \end{thm}

Theorem \ref{prva} is a special case of a result on
completeness of the set of polynomials which was obtained by the method of
shifting of arguments of the invariant polynomials on complex
semi-simple Lie algebras upon their restriction to normal
subalgebras \cite{MF1}. There is another proof by Bolsinov which is based on the
bi-Hamiltonian approach related to the compatibility of the
Lie-Poisson bracket \eqref{Lie-Poisson} with the Poisson bracket
\begin{equation}
\{f,g\}_A(M)=-\langle M, [\nabla f(M),\nabla g(M)]_A\rangle.
\label{**}
\end{equation}
The Lie algebra bracket $[M_1,M_2]_A=M_1AM_2-M_2AM_1$ is
compatible with the standard one in $\so(n)$ \cite{Bo}.

The system \eqref{E3} was also considered by Mishchenko.  He constructed
a set of integrals
\begin{equation}\label{MIS}
J_k=\sum_{p=1}^k\tr( B^{p-1}MB^{k-p}\Omega), \qquad k \ge 0,
\end{equation}
and proved their independence for $k=1,\dots,n-1$, $k\ne 2$ on
 generic adjoint orbits. This imply the complete integrability of the system for $n=4$
\cite{Mis}. The first integrals \eqref{MIS}  commute among each other (see
\cite{Di, Ra}) and they commute with the Manakov first integrals (see \cite{Ra}).

We consider the symmetric rigid body. In this case,
some of the eigenvalues of the mass tensor $B$ are equal, and the same is true for the
matrix $A=B^2$. We start from the orthogonal decomposition
\begin{equation}\label{v}
\so(n)=\so(n)_A\oplus\mathfrak v \cong \so(n_1)\oplus
\so(n_2)\oplus\dots\oplus \so(n_r)\oplus \mathfrak v,
\end{equation}
where $\so(n)_A=\{M\in \so(n)\, \vert\, MA=AM\}$ is the isotropy
subalgebra of $A$.

The Euler equations \eqref{E3} obey the Noether conservation law
$$
\pr_{\so(n)_A}M=const.
$$
By $\mathcal N$  the Noether first integrals are denoted. This is
a set of linear functions on $\so(n)_A$. From $\{\mathcal
L,\mathcal N\}=0$, we get that the polynomials in $\mathcal L$ are
$\Ad_{SO(n)_A}$--invariant, where $SO(n)_A\cong
SO(n_1)\times\dots\times SO(n_r)$ is a subgroup of $SO(n)$ with
the Lie algebra $\so(n)_A$.

The algebra of the Noether first integrals in general is not
commutative. Thus,  a natural approach toward the study of  symmetric rigid
body motions is provided by the noncommutative integration (see Nekhoroshev
\cite{N} and Mishchenko and Fomenko \cite{MF2}).

\begin{thm}\label{druga}
Consider the set $\mathcal L+\mathcal N$. It is a complete
noncommutative subset of functions on $\so(n)$ with respect to the
Lie--Poisson bracket (\ref{Lie-Poisson}). The equations of motion
of a symmetric rigid body \eqref{E3} are completely integrable in
the noncommutative sense.
\end{thm}

Bolsinov's proof of the above theorem, which  uses the compatibility of Poisson
brackets \eqref{Lie-Poisson} and \eqref{**} may be found in \cite{TF}, pages
241-244). Another proof, based on the symmetric pair decomposition of
$\gl(n)$, is presented in \cite{DGJ1}.

A natural question, related to the integrability of the
geodesic flows on a homogeneous space $SO(n)/SO(n)_A$ (see
\cite{BJ1, BJ3}), is the completeness of the restrictions of the
polynomials \eqref{integrals} to $\mathfrak v$
\begin{equation} {\mathcal L}_{\mathfrak v}=\{\tr(M+\lambda
A)^k\vert_\mathfrak v\,\, \vert \, \lambda\in \mathbb{R}, \,
k=1,2,\dots,n\}. \label{LV}
\end{equation}

Let us consider $\mathcal L_\mathfrak v$ within the algebra
$\mathbb{R}[\mathfrak v]^{SO(n)_A}$ of $\Ad_{SO(n)_A}$--invariant
polynomials on $\mathfrak v$ with respect to the restriction of
the Lie-Poisson bracket \eqref{Lie-Poisson}:
\begin{equation}
\{f,g\}_\mathfrak v(M)=-\langle M, [\nabla_\mathfrak v
f(M),\nabla_\mathfrak v g(M)] \rangle, \qquad  f,g \in
\R[\mathfrak v]^{SO(n)_A}. \label{THIMM}
\end{equation}
The set $\mathcal L_\mathfrak v$ is a commutative subset of
$\mathbb{R}[\mathfrak v]^{SO(n)_A}$. It is complete if in
$\mathcal L_\mathfrak v$ there are
\begin{equation}
\label{uslov1} N_\mathfrak v=\dim \mathfrak v-\frac12 \dim
{\mathcal O}_{SO(n)} (M)
\end{equation}
functionally independent polynomials, for a generic $M\in
\mathfrak v$, where ${\mathcal O}_{SO(n)}(M)$ is an adjoint orbit
of $SO(n)$ (see \cite{BJ1, BJ3}).

\begin{thm}\label{treca}
Consider the set ${\mathcal L}_\mathfrak v$. It is a complete
commutative subset of the algebra $\mathbb{R}[\mathfrak
v]^{SO(n)_A}$.
\end{thm}

Observe that the polynomials within $\mathbb{R}[\mathfrak
v]^{SO(n)_A}$ can be considered as $SO(n)$--invariant functions,
polynomially dependent on the momenta, on the cotangent bundle of
the homogeneous space $SO(n)/SO(n)_A$. The problem of construction
of complete commutative algebras of functions, polynomial in
momenta, within the algebra of $G$--invariant functions on
cotangent bundles of homogeneous spaces $T^*(G/H)$ is related to
the Mishchenko-Fomenko conjecture: noncommutative integrability
implies the usual Liouville integrability by means of first
integrals that belong to the same functional class as the
noncommutative first integrals \cite{MF2, TF}. This conjecture is
proved for $C^\infty$--smooth case for infinite-dimensional
algebras of first  integrals (see \cite{BJ2}). It is also proved
for the polynomial and analytic cases for finite-dimensional
algebras of first integrals (see \cite{Sa, Bo2}). For homogeneous
spaces $G/H$ there are several known constructions of complete
commutative $G$--invariant algebras (see \cite{BJ1, M, BJ3, MP,
Jo, Jo2}). However,  the general problem remains open.

Theorem \ref{treca} has been formulated as Theorem 4 in
\cite{DGJ1}. However, the first part of the proof presented there
needed to be completed, as it was noticed in \cite{DGJ2}. The
relation (29) of \cite{DGJ1} (the completeness of $\mathcal
L+\mathcal N$) holds on a Zariski open set $U\subset\so(n)$. In
\cite{DGJ2} it is shown that $\mathcal L+\mathcal N$ is complete
at $\mathfrak v$, i.e., $U\cap\mathfrak v\ne 0$, if and only if
$n_1,\dots,n_r \le 2$ (see page 1287, \cite{DGJ2}). Therefore, the
relation (36) of \cite{DGJ1} needs an additional argumentation.
Recently Mykytyuk provided a proof of Theorem \ref{treca} based on
bi-Hamiltonain methods (see \cite{M2}).

The aim of this note is to present  simple, direct proofs of
Theorems \ref{prva} and \ref{treca} for a family of matrices $A$
with multiple eigenvalues  (see Section 3). The proof is analogous
to the Mishchenko proof of the independence of integrals
\eqref{MIS} of the Euler equations \eqref{E3} by calculating their
gradients in a specific point of $\so(n)$ (see \cite{Mis}).

\section{Completeness of the integrals}

By expanding the integrals \eqref{integrals} in $\lambda$, using
the fact that the product of a symmetric and a skew-symmetric
matrix is traceless, we obtain
\begin{eqnarray*}
&&\tr((M+\lambda A)^2)=\tr(M^2)+\lambda^2\tr (A^2),\\
&&\tr((M+\lambda A)^3)=\lambda\tr (M^2A+MAM+AM^2)+\lambda^3\tr(A^3),\\
&&\tr((M+\lambda A)^4)=\tr(M^4)+\\
&&+\lambda^2\tr (M^2A^2+MAMA+MAAM+AMMA+AMAM+A^2M^2)+\lambda^4\tr(A^4),\\
&&\dots
\end{eqnarray*}

Apart from the leading terms that are constants, we get the
Manakov polynomial integrals in the form
\begin{eqnarray*}
&& p^k_{l}=\tr(M^{2k}A^l+(\text{all other permutations of a multiset with 2{\it k M}'s and {\it l  A}'s})),\\
&& k=1,2,\dots,[n/2], \qquad l=0,1,\dots,n-2k.
\end{eqnarray*}

Note that for $SO(n)_A$-invariant polynomial $f$ on $\mathfrak v$,
we have $\pr_{\so(n)_A}[M,\nabla_\mathfrak v f\vert_M]=0$ and
thus,  $\mathfrak v$-valued Hamiltonian vector field $X^\mathfrak
v_f=[M,\nabla_\mathfrak v f]$ is well defined. Since the gradient
$\nabla_\mathfrak v \tilde p^k_{l}$ of the restriction $ \tilde
p^k_l=p^k_{l}\vert_\mathfrak v $ is the projection $
\nabla_\mathfrak v \tilde p^k_{l}=\pr_\mathfrak v(\nabla
p^k_{l})$, the Hamiltonian vector field of $\tilde p^k_{l}$ is
given by
$$
X^{\mathfrak v}_{\tilde p^k_l}(M)=[M,\nabla_\mathfrak v \tilde
p^k_{l}\vert_M]=[M,\nabla p^k_{l}\vert_M]-[M,\pr_{\so(n)_A}\nabla
p^k_{l}\vert_M],\qquad M\in\mathfrak v.
$$

One can simply verify that the gradient $\nabla p^k_{l}\vert_M$ is
proportional to
\begin{eqnarray*}
P^k_{l}(M)=M^{2k-1}A^l+(\text{all other permutations of a multiset
with 2{\it k}-1 {\it M}'s and {\it l A}'s}),
\end{eqnarray*}
and, therefore,  the Hamiltonian vector field of $\tilde p^k_{l}$
is proportional to
\begin{equation}\label{HVF}
Q^k_l(M)=[M,P^k_l(M)]-[M,\pr_{\so(n)_A} P^k_l(M)]\in\mathfrak v,
\qquad M\in\mathfrak v.
\end{equation}

Define the vector subspaces
\begin{eqnarray*}
&&U_k=U_k(M)=\Span\{Q^k_{l}(M)\,\vert\,l=0,1,\dots,n-2k\},\\
&& U=U(M)=U_1+U_2+\dots+U_{[n/2]}.
\end{eqnarray*}

\subsection{Regular case}
Without loss of generality, suppose $ n_1\le n_2\le n_3\dots \le
n_r. $ If the condition
\begin{equation}
n_r \le n_1+\dots+n_{r-1}+1 \label{regular}
\end{equation}
is satisfied, then a generic element $M\in \mathfrak v$ is regular
element of $\so(n)$ and the required number of independent
integrals \eqref{uslov1} becomes
\begin{equation}
\label{uslov2} N_\mathfrak
v=N_{\so(n)}-\dim\so(n)_A=\frac12\left(\frac{n(n-1)}{2}+\left[\frac{n}{2}\right]\right)-\sum_{i=1}^r\frac{n_i(n_i-1)}{2}\,.
\end{equation}

First, we assume that the condition \eqref{regular} is satisfied.
Then a generic element $M\in \mathfrak v$ is regular element of
$\so(n)$ and the invariant polynomials
\begin{equation}\label{invarijante}
p^1_0,\dots,p^{[n/2]}_0
\end{equation}
are independent on $\mathfrak v$. Since their Hamiltonian vector
fields are zero, we need to prove
\begin{equation}\label{u1}
\dim U(M)=N_{\mathfrak v}-[n/2]=N_{\so(n)}-\dim\so(n)_A-[n/2],
\end{equation}
for a generic $M\in\so(n)$. Note that the total number of
polynomials $p^k_{l}$ is
\begin{equation*}\label{broj}
 (n-1)+(n-3)+\dots+(n+1-2[n/2])=N_{\so(n)}.
\end{equation*}

Since we deal with polynomials, it is enough to prove the
conditions \eqref{u1} in one point $M_0$ only. In \cite{Mis},
Mishchenko considered gradients of integrals \eqref{MIS} at the
point $\sum_{i>1} E_1\wedge E_i+\sum_{k>1} E_{2k-1}\wedge E_{2k}.$
Similarly, here we consider the gradients at the point
\begin{equation}\label{M0}
M_0=\sum_{i} E_i\wedge E_{i+1},
\end{equation}
where, using the condition \eqref{regular}, we take a permutation
of the components $a_i$ of the matrix $A$, such that
$M_0\in\mathfrak v$. Note that a permutation of the diagonal
elements of the matrix $A$ leads to the equivalent problem. We
will keep the same symbol for the new matrix $A$.

Consider vector subspaces (see Figure 1)
\begin{eqnarray*}
&& W_k=\Span\{E_{1}\wedge E_{2k},E_{2}\wedge
E_{2k+1},\dots \}, \qquad k=1,2,\dots,[n/2]\\
&& V_k=\Span\{E_{1}\wedge E_{2k+1},E_{2}\wedge E_{2k+2},\dots\},\\
&&\mathfrak v_k=V_k\cap \mathfrak v, \qquad \qquad\qquad
\qquad\qquad \qquad\qquad k=1,2,\dots,[n/2]
\end{eqnarray*}
($V_{[n/2]}=0$ if $n$ is even and $V_{[n/2]}=\Span\{E_1\wedge
E_n\}$ if $n$ is odd).

\begin{figure}[ht]
\includegraphics[width=110mm, height=75mm]{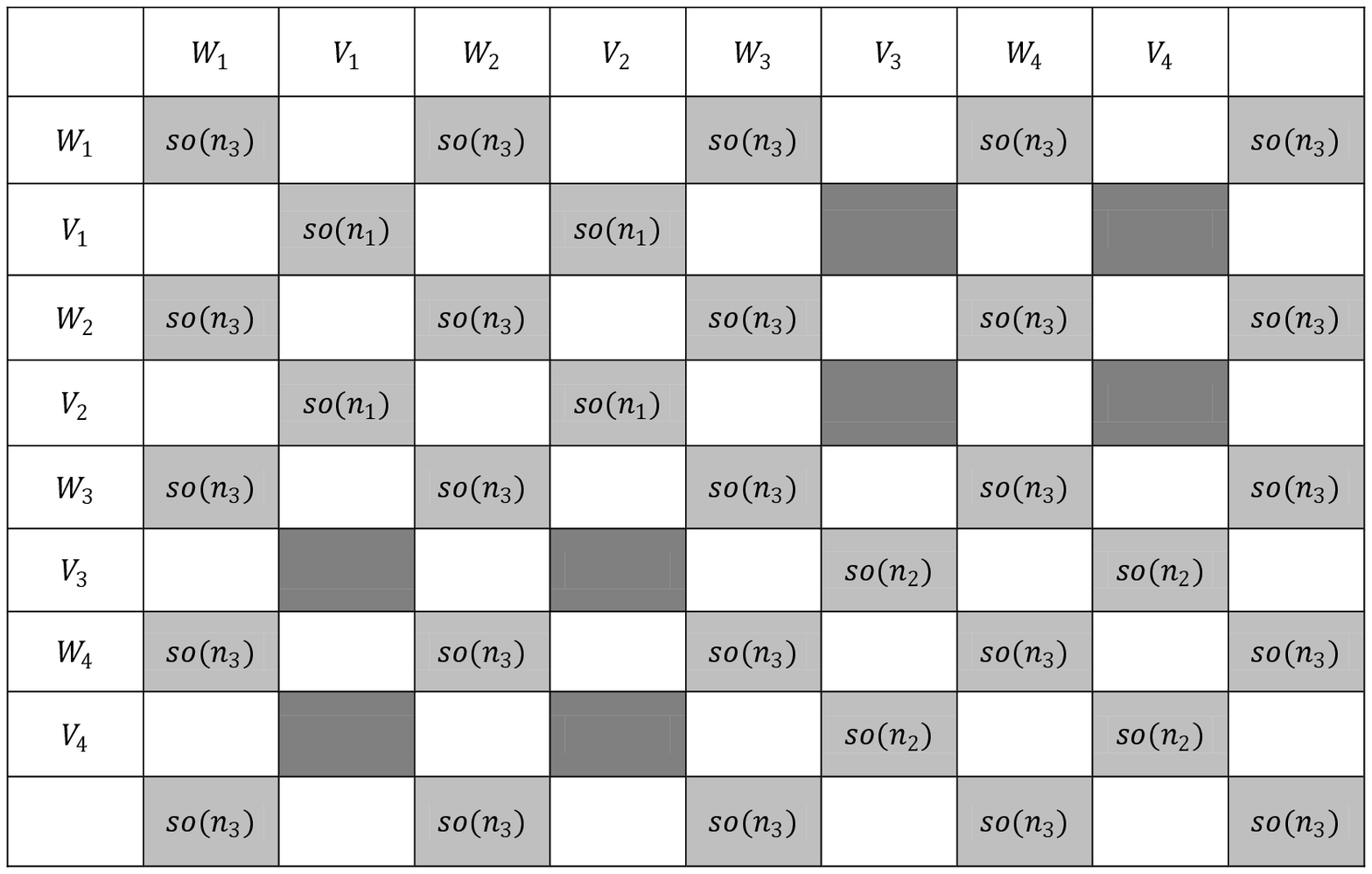}
\caption{An illustration for the decomposition
$\so(9)=\so(2)\oplus \so(2)\oplus \so(5)\oplus \mathfrak v$. We
have $\mathfrak v=W_1\oplus \mathfrak v_1\oplus W_2\oplus
\mathfrak v_2\oplus W_3\oplus \mathfrak v_3\oplus
W_4\oplus\mathfrak v_4$, $\dim\mathfrak v_1=1$, $\dim\mathfrak
v_2=2$, $\dim\mathfrak v_3=1$, and $\dim\mathfrak v_4=0$.}
\end{figure}

From the form of expressions of $P^k_l$ at the point $M_0$ and
\eqref{HVF} it follows:
\begin{eqnarray*}
P^k_l(M_0) \in W_1 \oplus \dots \oplus W_k, \qquad Q^k_l(M_0) \in
\mathfrak v_1 \oplus \dots \oplus \mathfrak v_k.
\end{eqnarray*}

Therefore,
\begin{equation}
U_k(M_0)\subset \mathfrak v_1\oplus \mathfrak v_2\oplus
\dots\oplus \mathfrak v_k. \label{uk2}
\end{equation}

For a collection of numbers $(n_1,\dots,n_r)$, $n_1\le n_2\le
\dots\le n_r$, $n_r \le n_1+\dots+n_{r-1}+1$, in addition, we
suppose the following assumption:

\begin{as}
There exists a permutation of diagonal elements of the matrix $A$,
such that all $n_i$ elements equal to $\alpha_i$ are rearranged to
occupy the places with indexes $j_i,j_i+2,\dots,j_i+2(n_i-1)$,
$i=1,2,\dots,r$. Here the index $j_i$ denotes the first appearance
of an element equal to $\alpha_i$.
\end{as}

If the Assumption 1 is satisfied, then
 $ \so(n)_A$ is a subset of $V_1\oplus \dots \oplus V_{[n/2]}$, while the gradients
$P^k_l(M_0)$ belong to $W_1\oplus\dots\oplus W_k\subset\mathfrak
v$. Therefore, $\pr_{\so(n)_A} P^k_l(M_0)=0$ and
$Q^k_l(M_0)=[M_0,P^k_l(M_0)]$. Furthermore,
$$
\dim\mathfrak v_k=\dim V_k-\delta_k=(n-2k)-\delta_k, \qquad
\delta_k=\dim\so(n)_A\cap V_k,
$$
where
\begin{eqnarray*}
&&\delta_k=\delta_{k,1}+\dots+\delta_{k,r}, \qquad
\delta_{k,i}=\dim \so(n_i)\cap V_k\\
&&\delta_{k,i}=n_i-k, \quad k=1,\dots,n_i-1, \quad \delta_{k,i}=0,
\quad k\ge n_i.
\end{eqnarray*}

\begin{exm}{\rm
If $n=9$, $r=3$ $n_1=2, n_2=2,n_3=5$, starting from
$$
A=\diag(\alpha_1,\alpha_1,\alpha_2,\alpha_2,\alpha_3,\alpha_3,\alpha_3,\alpha_3,\alpha_3),
$$
a required permutation is (see Figure 1)
$$
A=\diag(\alpha_3,\alpha_1,\alpha_3,\alpha_1,\alpha_3,\alpha_2,\alpha_3,\alpha_2,\alpha_3).
$$
On the other hand, the choice $r=3$, $n_1=n_2=n_3=3$ does not
satisfy the Assumption 1. This is the only example for $n=9$,
which does not satisfy the Assumption 1.}
\end{exm}

The components $S^k_{l,i}$ of the projection of $Q^k_{l}(M_0)$ to
$V_k$ can be easily calculated:
\begin{eqnarray*}
&& S^1_{1,i}=\langle Q^1_1,E_i\wedge E_{i+2}\rangle=a_{i+2}-a_{i},\\
&& S^1_{2,i}=\langle Q^1_{2},E_i \wedge E_{i+2}\rangle=
(a_{i+2}-a_i)(a_i+a_{i+1}+a_{i+2}),\\
&& S^1_{3,i}=\langle Q^1_{3},E_i \wedge E_{i+2}\rangle=(a_{i+2}-a_{i})(a_ia_{i+1}+a_ia_{i+2}+a_{i+1}a_{i+2}+a_i^2+a_{i+1}^2+a_{i+2}^2),\\
&& \dots
\end{eqnarray*}
\begin{eqnarray*}
&& S^2_{1,i}=\langle Q^2_{1},E_i\wedge E_{i+4}\rangle=a_{i+4}-a_{i},\\
&& S^2_{2,i}=\langle Q^2_{2},E_i \wedge E_{i+4}\rangle=(a_{i+4}-a_i)(a_i+a_{i+1}+a_{i+2}+a_{i+3}+a_{i+4}),\\
&& S^2_{3,i}=\langle Q^2_{3},E_i \wedge E_{i+4}\rangle=(a_{i+4}-a_{i})\sum_{i \le i_1 \le i_i \le i+4} a_{i_1}a_{i_2},\\
&& \dots
\end{eqnarray*}
\begin{eqnarray*}
&& S^3_{1,i}=\langle Q^3_{1},E_i\wedge E_{i+6}\rangle=a_{i+6}-a_{i},\\
&& S^3_{2,i}=\langle Q^3_{2},E_i \wedge E_{i+6}\rangle=(a_{i+6}-a_i)(a_i+a_{i+1}+a_{i+2}+a_{i+3}+a_{i+4}+a_{i+5}+a_{i+6}),\\
&& S^3_{3,i}=\langle Q^3_{3},E_i \wedge E_{i+6}\rangle=(a_{i+6}-a_{i})\sum_{i \le i_1 \le i_i \le i+6} a_{i_1}a_{i_2},\\
&& \dots
\end{eqnarray*}

Therefore, for any condition
$$
a_{j_i}=a_{j_i+2}=a_{j_i+4}=\dots=a_{j_i+2(n_i-1)}=\alpha_i,
$$
$\delta_{k,i}$ columns $j_i,j_i+2,j_i+4,\dots,j_i+2(n_i-k-1)$ of a
$(n-2k)\times(n-2k)$-matrix $S^k$ are equal to zero
\begin{eqnarray*}
&& S^k_{1,j_i}=S^k_{2,j_i}=\dots=S^k_{n-2k,j_i}=0, \\
&& S^k_{1,j_i+2}=S^k_{2,j_i+2}=\dots=S^k_{n-2k,j_i+2}=0, \\
&& \dots\\
&&
S^k_{1,j_i+2(n_i-k-1)}=S^k_{2,j_i+2(n_i-k-1)}=\dots=S^k_{n-2k,j_i+2(n_i-k-1)}=0.
\end{eqnarray*}
Also, the non zero columns of  $S^k$ are independent. Hence, the
rank of $S^k$ is equal to $n-2k-\delta_k$ and
\begin{equation}\label{LEM2}
\pr_{\mathfrak v_k}(U_k(M_0))=\mathfrak v_k.
\end{equation}

Therefore, the relation \eqref{uk2} becomes an equality, and we
obtain the condition \eqref{u1}:
\begin{eqnarray*}
\dim U(M_0) &=& \sum_{k=1}^{[n/2]} \dim
V_k-\sum_{i=1}^r\sum_{k=1}^{n_i}(n_i-k)\\
&=& N_{\so(n)}-[n/2]-\sum_{i=1}^r\dim \so(n_i)=N_\mathfrak
v-[n/2].
\end{eqnarray*}

Besides, as a bi-product, from the above analysis we get that
\begin{equation}\label{nezavisni}
\mathcal L_\mathfrak v^0=\{p^k_{l}\vert_\mathfrak v
\quad\vert\quad k=1,2,\dots,[n/2], \quad
l=0,1,\dots,n-2k-\delta_k\}
\end{equation}
is a complete set of independent integrals.

We verified directly that the polynomials \eqref{nezavisni} are
independent even if the Assumption 1 does not hold in several
cases. However, the proofs are not so elegant and we feel that in
those cases, it is more natural to use a bi-Hamiltonian approach,
as it was done by Mykytyuk \cite{M2}.

\begin{exm}{\rm
For the case considered in Example 1, the restrictions of
\begin{equation*}
p^1_0,    \,  p_1^1,    p^2_0,\, p^2_1,  \, p^2_2,\,  p^3_0,
\,p^{3}_1, \, p^4_0
\end{equation*}
to $\mathfrak v$ provide a  complete commutative set on
$\R[\mathfrak v]^{SO(2)\times SO(2)\times SO(5)}$. }\end{exm}

\subsection{Singular case}
Next, we consider the case when
\begin{equation}\label{regular2}
n_r = n_1+\dots+n_{r-1}+s, \qquad s>1.
\end{equation}

Let $n'=n-s$ and let $\so(n')=so(n')_{A'}\oplus \mathfrak v'$ be
the orthogonal decomposition, where $A'$ is the $n'\times
n'$-diagonal matrix obtained from $A$ by removing $s$ components
equal to $\alpha_r$. Then
\begin{equation}\label{vv}
\so(n')_{A'} \cong \so(n_1)\oplus \so(n_2)\oplus\dots\oplus
\so(n_{r-1})\oplus \so(n_r-s),
\end{equation}
and  the condition \eqref{regular} is satisfied for $\so(n')$ and
$A'$. Thus, if the Assumption 1 for $(n_1,\dots,n_{r-1},n_r-s)$ is
satisfied, the corresponding set $\mathcal L_{\mathfrak v'}^0$
given by \eqref{nezavisni} is a complete commutative set on
$\mathfrak v'$. On the other hand, from the second part of the
proof of Theorem 4 \cite{DGJ1} we have
$$
N_{\mathfrak v}=N_{\mathfrak v'}.
$$
Thus, the set of polynomials $\mathcal L_{\mathfrak v'}^0$ is a
complete commutative set of polynomials on $\mathfrak v$, where we
instead the restrictions $p^k_l\vert_{\mathfrak v'}$ we take the
restrictions $p^k_l\vert_{\mathfrak v}$.

\

As a result, we proved the completeness of Manakov integrals for
homogeneous spaces $SO(n)/SO(n_1)\times\dots\times SO(n_r)$, for
collections of numbers $n_1 \le n_2 \le \dots \le n_r$ described
by

\begin{itemize}
\item{} $n_r \le n_1+\dots+n_{r-1}+1$ and $(n_1,\dots,n_r)$
satisfy the Assumption 1;

\item{}  $n_r = n_1+\dots+n_{r-1}+s$, $s>1$ and
$(n_1,\dots,n_{r-1},n_r-s)$ satisfy the Assumption 1.
\end{itemize}

In particular, if all eigenvalues of $A$ are distinct then
$\mathfrak v=\so(n)$ and we obtain that all integrals $p^k_l$ are
functionally independent. In that case, the above consideration
can be considered as a simplified version of the
Mishchenko--Fomenko theorem on the completeness of polynomials on
normal subalgebras of complex semi-simple Lie algebras \cite{MF1}.

\subsection*{Acknowledgments}
The research was supported by the Serbian Ministry of Science
Project 174020 Geometry and Topology of Manifolds, Classical
Mechanics, and Integrable Dynamical Systems. We are grateful to S.
S. Nikolaenko for the correction of few misprints.

\

\end{document}